\documentclass[sigconf,edbt]{acmart-edbt2021}

\def\BibTeX{{\rm B\kern-.05em{\sc i\kern-.025em b}\kern-.08em
    T\kern-.1667em\lower.7ex\hbox{E}\kern-.125emX}}

\usepackage{booktabs} 

\usepackage{ascmac}

\newcommand{\miscelaV}{{\sc Miscela-V}}
\newcommand{\miscela}{{\sc Miscela}}

\setcopyright{rightsretained}

\acmDOI{}

\acmISBN{978-3-89318-084-4}

\acmConference[EDBT 2021]{24th International Conference on Extending Database Technology (EDBT)}{March 23-26, 2021}{Nicosia, Cyprus} 
\acmYear{2021}

\begin{document}
\title{Smart City Data Analysis via Visualization of Correlated Attribute Patterns}

\author{Yuya Sasaki$^{1*}$, Keizo Hori$^{1*}$,  Daiki Nishihara$^{1*}$, Sora Ohashi$^{1*}$, Yusuke Wakuta$^{1*}$, Kei Harada$^{1}$, Makoto Onizuka$^{1}$, Yuki Arase$^{1}$, Shinji Shimojo$^{2}$, Kenji Doi$^{3}$, He Hongdi$^{4}$, Zhong-Ren Peng$^{5}$}
\thanks{$^*$ These authors contributed equally. Yuya Sasaki is the corresponding author}
\affiliation{%
  \institution{$^1$Graduate School of Information Science and Technology, Osaka University, Suita, Japan,
  $^2$Cyber media center, Osaka University, Suita, Japan,
  $^3$Graduate School of Engineering, Osaka University, Suita, Japan,
  $^4$Center for Intelligent Transportation Systems and Unmanned Aerial Systems Applications Research, Shianghai Jiao Tong University, Shanghai, China, and
  $^5$International Center for Adaptation Planning and Design, University of Florida, Gainesville, USA.\\
 \{sasaki, hori.keiso, nishihara.daiki, ohashi.sora, wakuta.yusuke,  harada.kei, onizuka, arase\}@ist.osaka-u.ac.jp, shimojo@cmc.osaka-u.ac.jp, doi@civil.eng.osaka-u.ac.jp, hongdihe@sjtu.edu.cn, zpeng@ufl.edu
 }}
\renewcommand{\shortauthors}{}

\begin{abstract}
Urban conditions are monitored by a wide variety of sensors that measure several attributes, such as temperature and traffic volume. The correlations of sensors help to analyze and understand the urban conditions accurately.
The correlated attribute pattern (CAP) mining discovers correlations among multiple attributes from the sets of sensors spatially close to each other and temporally correlated in their measurements. In this paper, we develop a visualization system for CAP mining and demonstrate analysis of smart city data. Our visualization system supports an intuitive understanding of mining results via sensor locations on maps and temporal changes of their measurements. In our demonstration scenarios, we provide four smart city datasets collected from China and Santander, Spain. We demonstrate that our system helps interactive analysis of smart city data.

\end{abstract}

%
%



\maketitle

\section{introduction \label{sec:introduction}}
Many cities have started smart city initiatives and installed a wide variety of sensors that measure several attributes, such as traffic volume and temperature.
The collected data from smart cities is used for continuously and cooperatively monitoring urban conditions, such as the distribution of air pollution, the transition of traffic volume, and the change of citizen activity.
Researchers and municipalities analyze smart city data and make a decision for urban planning.
For example, environmental researchers in Shanghai Jiao Tong university analyze the relationships between traffic and air pollution \cite{wang2020regional,jiang2017transport}.  
Santander, Spain monitors the traffic volumes within the city and informs people of the real-time traffic information~\cite{smartsantander}.
They work on obtaining useful patterns in cities by using database and data mining techniques.

Smart city data has spatial and temporal information.
For analysing spatio-temporal data, we proposed {\it correlated attribute pattern (CAP) mining} \cite{harada2019miscela,harada2020miscela}.
CAP mining aims to find correlated attributes of sensors that are spatially close to each other and whose measurements temporally co-evolve.
We developed an efficient algorithm for CAP mining, called \miscela{} and presented that the correlated attribute patterns can discover useful knowledge from smart city data.
We show an example that illustrates the effectiveness of CAP mining.

\begin{figure}[t]
  \begin{tabular}{c}
    \begin{minipage}{0.35\hsize}
    \centering
    \includegraphics[clip, width=\linewidth]{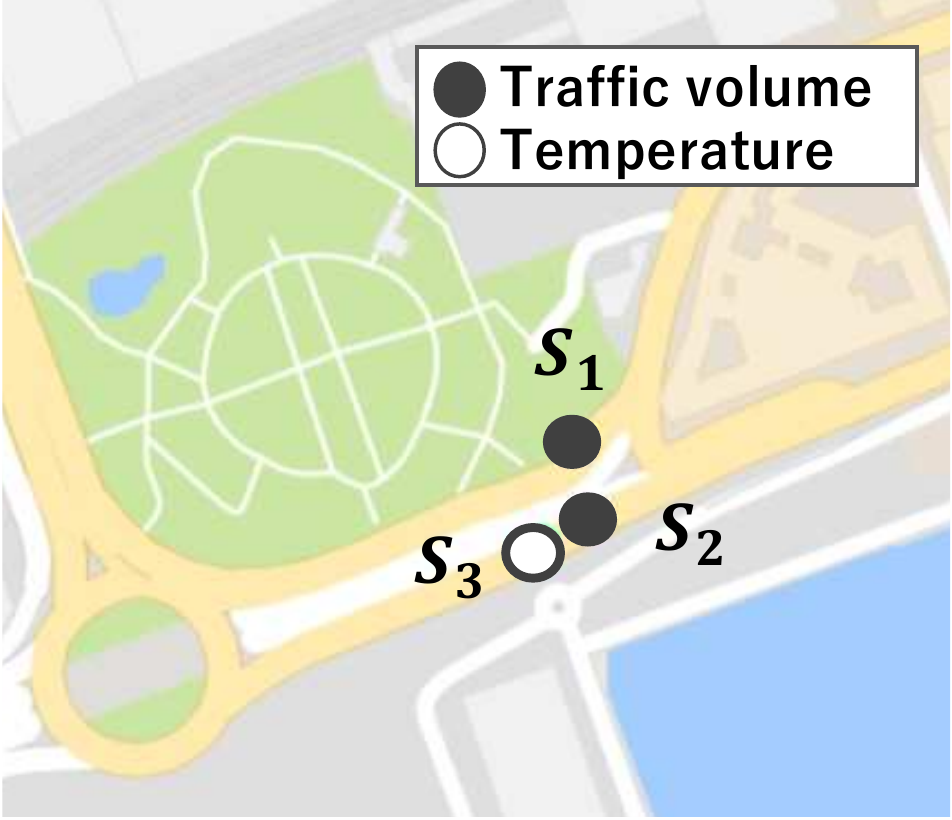}
    \hspace{10cm}(a) Sensor location
    \end{minipage}
    \begin{minipage}{0.65\hsize}
    \centering
    \includegraphics[clip, width=\linewidth, height=2.64cm]{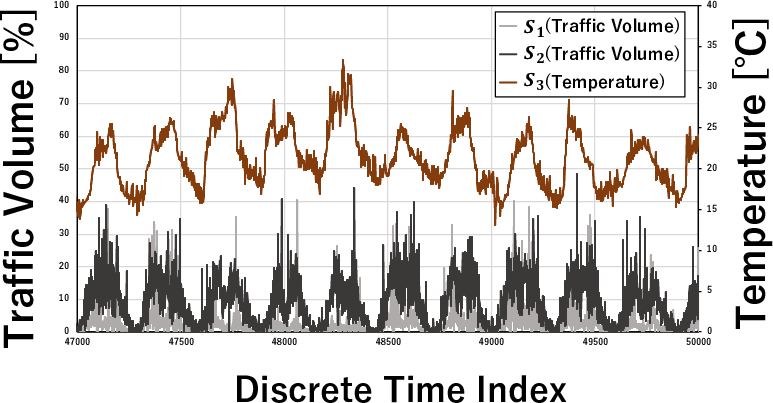}
    \hspace{10cm}(b) Correlation
    \end{minipage}
  \end{tabular}
  \caption{The correlation between traffic volume and temperature in Santander \cite{harada2019miscela}}
  \label{fig:CAP1}
\end{figure}

\begin{example}
Figure~\ref{fig:CAP1} shows locations of three sensors $s_1$, $s_2$, and $s_3$ in Santander and these measured values. $s_{1}$ and $s_{2}$ measure traffic volume and $s_{3}$ measures temperature.
These sensors are spatially close to each other, and the measurements of them co-evolve frequently (i.e., change the values simultaneously).
The CAP mining can discover correlated patterns among traffic volume and temperature measured by the three sensors.
Municipalities can understand that traffic behavior in the area is correlated to temperature from the CAP.
\end{example}

\smallskip
\noindent
{\bf Contribution:} In this paper, we develop a visualization system for CAP mining, called \miscelaV, to support an intuitive analysis of smart city data. 
\miscelaV{} has the following characteristics:
\begin{itemize}
    \item \miscelaV{} natively supports CAP mining with user-specified parameters.
    \item \miscelaV{} visualizes sensor locations on a map and temporal changes of sensor measurements.
    \item \miscelaV{} caches results of CAP mining and reuses the cached results for efficient interactive analysis.
\end{itemize}


Our system supports intuitive understanding of analytic results via visualization.
We demonstrate an analysis of smart city data by using our system.
We use two different scale datasets: Santander (i.e., city size) and China (i.e., country size).
Our system is effective for any space and time scales such as daily city-scale and minutely country-scale datasets. 
For further investigation, we open our source codes\footnote{\url{https://github.com/OnizukaLab/MISCELA-v}}.

This work is a collaborated work with researchers in database, environmental, and urban science fields, so we validated that \miscelaV{} is effective for environmental and urban science studies. 
Through the demonstration of \miscelaV, we expect that \miscelaV{} helps researchers in more other fields for accelerating their analysis.


\smallskip
\noindent
{\bf Related systems:} There are several systems for visualizing spatio-temporal data (e.g.,  \cite{xiufeng2020vap,hengl2015plotkml,andrienko2013visual}).
Some systems support spatial-temporal pattern mining but no systems support CAP mining.
The novelty of our system is that it focuses on CAP mining with efficient interactive analysis.

\smallskip
\noindent
{\bf Organization:}
The rest of this paper is organized as follows.
We explain the CAP mining and the CAP mining method \miscela{} in Section~\ref{sec:preliminaries} as preliminaries.
Then, we present a visualization system \miscelaV{} in Section~\ref{sec:system}.
After that, we show our demonstration plan in Section~\ref{sec:demo}, followed by the conclusion in Section~\ref{sec:conclusion}.

\section{Preliminaries}
\label{sec:preliminaries}

We explain CAP mining and \miscela{} as preliminaries.

\subsection{CAP mining}

We consider a sensor set in a geographical region.
Each sensor has longitude and latitude as spatial information.
It measures a specific attribute, such as temperature, traffic volume, and PM2.5.
Each sensor is synchronized, that is, it measures its sensor value at a certain interval.
We define that measurements are co-evolved if they increase/decrease at the same timestamp.

The CAP mining aims for discovering spatially and temporally correlated environmental properties such that multiple sensors measure those attributes that satisfy the following conditions:
(1) the set of sensors are located at spatially close locations to each other,
(2) the measurements of the sensors co-evolve frequently, and
(3) the set of attributes measured by the sensors includes multiple attributes.
The CAP mining restricts the correlation between different attributes to support diversified analysis of smart cities. This restriction can be easily removed.

CAP mining has several parameters for obtaining CAPs that users want.
We here summarize parameters and their impacts on the number of CAPs to be discovered. 
\begin{itemize}
    \item Evolving rate $\varepsilon$: The CAP mining removes slight changes of measurements by specifying $\varepsilon$. If the amount of changes from the previous timestamp is smaller than  $\varepsilon$, the timestamps are evaluated as that the measurements do not change.
    If $\varepsilon$ is large, sensors likely co-evolve, so the number of CAPs likely becomes large. 
    \item Distance threshold $\eta$: $\eta$ gives a criterion of close sensors. If a distance between the two sensors is less than $\eta$, we define that they are close. If $\eta$ is large, many sensors are spatially close to each other. 
    \item The maximum number of CAP attributes $\mu$: $\mu$ restricts the number of attributes in CAPs. The CAP mining discovers correlations among not larger than $\mu$ attributes. 
    \item The minimum support $\psi$: $\psi$ is the minimum support. If measurements of two sensors co-evolve at more than $\psi$ timestamps, they are co-evolving sensors. If $\psi$ is small, many sensors become co-evolving sensors, and thus the number of CAPs likely becomes large.
\end{itemize}
Since the sensitivity of parameters depends on datasets, it is necessary to support interactive analysis.
Please see more detailed definitions in~\cite{harada2019miscela}.

\begin{figure*}[!t]
    \centering
    \includegraphics[clip, width=0.9\linewidth]{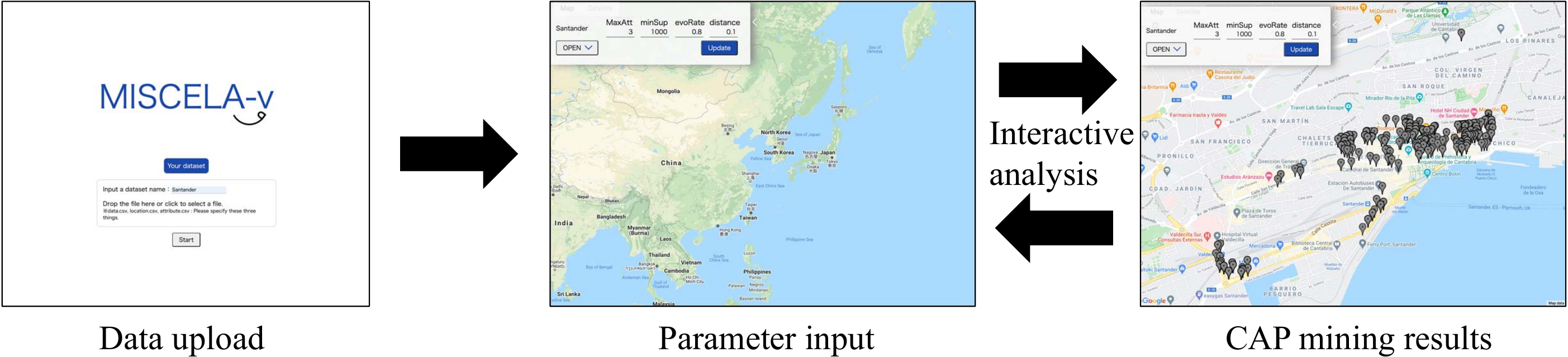}
    \caption{An overview of \miscelaV}
    \label{fig:miscelav_overview}
\end{figure*}

\begin{figure*}[!t]
    \centering
    \includegraphics[clip, width=0.85\linewidth]{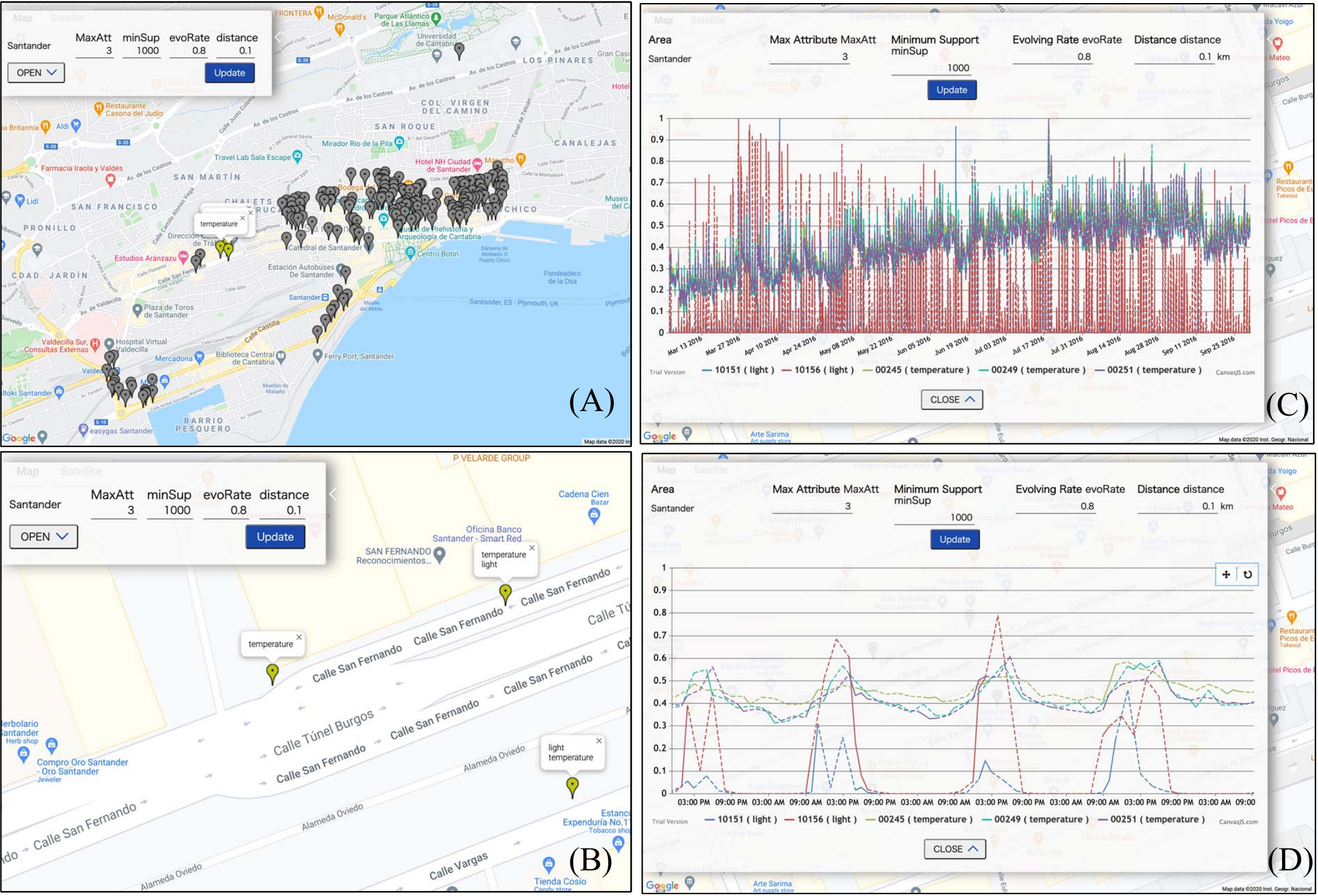}
    \caption{Visualization of CAP mining results}
    \label{fig:visualization}
\end{figure*}

\subsection{MISCELA: an efficient algorithm for CAP mining}
\miscela{} supports efficient computation for CAP mining, which comprises the following four steps.

\begin{enumerate}
    \item \textbf{Linear segmentation}:
    We filter uninteresting data fluctuation by applying a linear segmentation algorithm to time series data. 
    \item \textbf{Extracting evolving timestamps}:
    We extract evolving timestamps in the measurements of all sensors by using the given evolving rate $\varepsilon$.
    \item \textbf{Discovering spatially connected sets of sensors}:
    Since CAPs are discovered only from spatially connected sets,
    we divide a given sensor set into spatially close sensors to restrict the search space.
    \item \textbf{CAP search}:
    For each set of spatially close sensors, we search for CAPs.
    We recursively conduct the CAP search with gradually expanding spatially close sensors according to a tree structure for CAP mining.
\end{enumerate}

Please see more detailed and precise procedures in~\cite{harada2019miscela}.


\section{MISCELA-V: Visualizing System}
\label{sec:system}

We present our visualization system, which we call \miscelaV.
The purposes of \miscelaV{} is (1) to easily find CAPs in users' datasets, (2) to visually understand the CAPs, and (3) to efficiently support interactive analysis.
First, \miscelaV{} natively supports CAP mining.
It visualizes locations of sensors and changes of their measurements to understands reasons why these attributes are correlated.
In addition, since \miscela{} may take a large execution depending on their parameters, it has a caching mechanism for efficient interactive CAP mining.

\subsection{System overview}

Figure \ref{fig:miscelav_overview} shows an overview of \miscelaV.
\miscelaV{} has three main processes to visualize CAP mining results.
First, we upload datasets to the system.
Then, we input parameters of CAP mining for obtaining appropriate results.
Finally, we can see the CAP results on a map and the temporal behaviors of their measurements.
Since our system supports interactive analysis, data and CAPs are stored in databases.
Users can easily change parameters to check CAPs in different parameters.
If users specify the parameters of CAPs stored in databases, we can immediately see CAPs without processing \miscela.

Figure \ref{fig:visualization} shows a visualization of sensor locations and temporal measurements.
Figures (A) and (B) show sensor locations, and three sensors are highlighted. 
When we click a sensor in the map, sensors are highlighted if their measurements are correlated to measurements of the clicked sensor.
In addition, we can see the attributes of correlated sensors.
Figures (C) and (D) show temporal behaviors of measurements, which we can zoom in and zoom out.
In (D), you can see that three measurements frequently increase/decrease together. 
Our visualization helps to intuitively understand correlations among sensors.

\subsection{Data upload}

We can easily upload our datasets via a user interface that provides two ways of data upload:  drag-and-drop and selecting files from finder. 
For uploading datasets, we need to prepare three files; data.csv, location.csv, and attribute.csv.
data.csv lists the set of measurements at all timestamps. We note that timestamps must be the same time intervals, and sensor values are null if the sensors do not have the sensor values at timestamps.
location.csv lists the sensor information; identifier, attribute, and location.
attribute.csv lists all attributes in the datasets.
Each file should have the following formats:

\begin{itembox}[l]{data.csv}
id,attribute,time,data\\
00000,temperature,2016-03-01 00:00:00,null\\
00000,temperature,2016-03-01 01:00:00,9.87\\
$\cdots$
\end{itembox}

\begin{itembox}[l]{location.csv}
id,attribute,lat,lon\\
00000,temperature,43.46192,-3.80176\\
00001,temperature,43.46212,-3.79979\\
$\cdots$
\end{itembox}

\begin{itembox}[l]{attribute.csv}
temperature\\
light\\
$\cdots$
\end{itembox}

The data.csv might be very large. For scalably uploading large datasets, we divide the file into 10,000 lines and send each divided set to our system.
Each dataset is stored in databases, and thus we can use the dataset without re-uploading by specifying the dataset name.


\subsection{Caching mechanism}
\miscela{} may take a long time for finding CAPs depending on data and user-specified parameters.
For efficient interactive analysis, \miscelaV{} caches CAP mining results and reuses the cached results if users specify the same parameter setting.
This caching mechanism accelerates the analytic process and reduces the computational costs when the front end receives multiple requests at the same time.

We store the name of the dataset, parameters, and CAPs (i.e., a set of sets of sensors) to the database.
Before computing CAPs by \miscela, our system searches for CAPs with the same parameters and the name of the dataset from the database.
Since interactive analysis could input the same parameters to compare results repeatedly, the caching mechanism supports more efficient data analysis.



\subsection{Implementation}
We use MongoDB as database management systems and django as API servers. \miscela{} is implemented by Python, and a map visualization is implemented by JavaScript, jQuery, and Google Map API. 
\miscela{} returns a set of sets of sensors as CAPs that might include many sensors (or empty), and its format is JSON.
Since RDBMS is not suitable for \miscela{} outputs, we select MongoDB to store datasets and CAP results. 
Since we design that these components are connected by APIs, we can modify each component individually.



\section{Demonstration Plan}
\label{sec:demo}
For \miscelaV{} demonstrations, we use smart city data in Santander and China, as a case study.
We will introduce the system architecture, the analytic process, and how to use our system to find knowledge. 
Attendees can interact with our system to perform analysis using the data.
For example, since \miscelaV{} can show temporal changes of sensors' measurements, we can analyze the difference of measurements before/after COVID-19.
The attendees will interactively discover CAPs of smart city data.

The attendees can use the following datasets\footnote{We consider sensors with different attributes as different sensors even if they are located at the same location.}: 
\begin{itemize}
\item {\bf Santander} includes 552 sensors in Stantander, Spain from  2016 March 1st to September 30th. The number of records is 2{,}329{,}936.
Attributes are temperature, light, sound, traffic volume, and humidity.
\item {\bf China6} includes 9{,}438 sensors in China from 2016 September 1st to 2018 October 31st. {The number of records is 6{,}889{,}740.}
Attributes are PM2.5, SO2, NO2, CO, and O3.
\item {\bf China13} includes 4{,}810 sensors. The period is the same as China6. The number of records is 3{,}511{,}300. Attributes are additionally included in temperature, humidity, air pressure, daylight, rainfall percentage, rain volume, and wind speed.
\item {\bf COVID-19} includes 12 sensors in Shanghai and Guangzhou, China from  2020 January 1st to June 30th. The number of records is 52{,}261. Attributes are PM2.5, PM10, SO2, NO2, CO, and O3. This data includes the period after and before spreading COVID-19. 
\end{itemize}

We plan to demonstrate the following case studies.

\noindent
{\bf Interactive analysis: }
In this demonstration, we first provide interactive analysis to upload datasets, input parameters, and view CAP results.
Attendees can freely use our system and try to find interesting patterns in our datasets.
First, attendees set the parameters for finding CAPs and see the visualization of the results.
Second, the attendees can investigate why the CAPs are discovered by visualizing the temporal behavior of measurements of sensors.
Since our system highlights sensors that are correlated, they can understand what sensors are correlated intuitively.

\noindent
{\bf Santander dataset: a single city data analysis: }
This scenario aims to find interesting knowledge within Santander. 
Attendees will find interesting CAPs from Santander datasets and investigate the results via visualization.
For example, we can find correlated patterns among temperatures and traffic volumes and among light and temperature.

\noindent
{\bf China dataset: multiple cities data analysis: }
This scenario aims to find interesting knowledge among many cities in China.
In particular, attendees can intuitively understand that two sensors are correlated even if they are distant from each other.
Furthermore, sensors are not correlated if two sensors are vertically (north and south) close to each other, but if sensors are horizontally (east and west) close, they are correlated.
These are often caused by wind directions. We can understand that wind directions affect to air quality from the CAPs.
Our system supports for understanding reasons why sensors are correlated and not correlated.

\begin{figure}[t]
  \begin{tabular}{c}
    \begin{minipage}{0.5\hsize}
    \centering
    \includegraphics[clip, width=\linewidth]{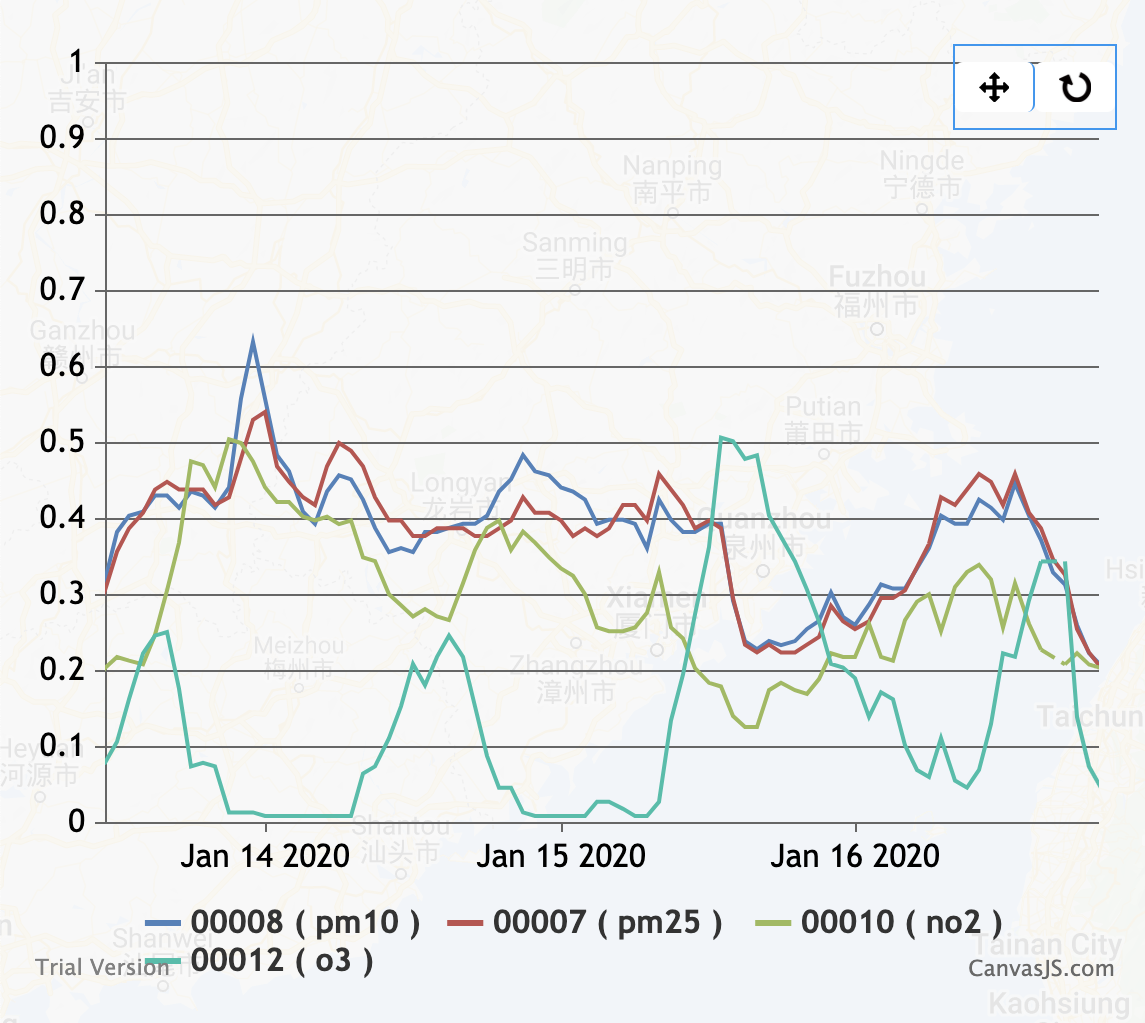}
    \hspace{10cm}(a) Before
    \end{minipage}
    \begin{minipage}{0.5\hsize}
    \centering
    \includegraphics[clip, width=\linewidth]{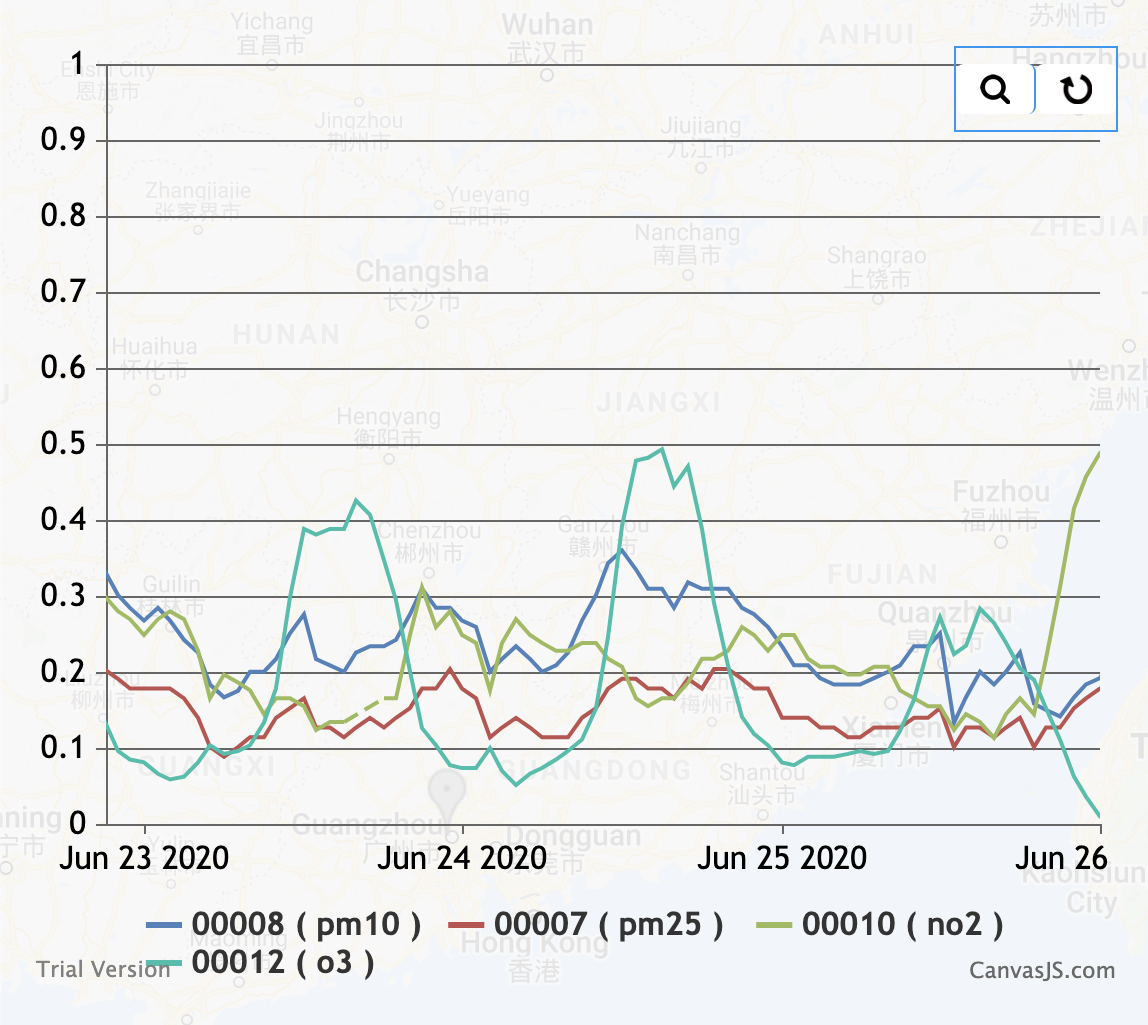}
    \hspace{10cm}(b) After
    \end{minipage}
  \end{tabular}
  \caption{An example of correlation pattern changes before/after spreading COVID19}
  \label{fig:covid}
\end{figure}

\noindent
{\bf COVID-19 analysis: }
COVID-19 dataset includes the period before and after spreading COVID-19.
Attendees can know that levels of air pollution change due to spreading COVID-19.
Figure~\ref{fig:covid} shows the correlation patterns before and after COVID-19. 
From these results, we can visually understand that our activity changes affect not only the amounts of air pollutants but also their correlation patterns.


\section{Conclusion}
\label{sec:conclusion}
In this paper, we introduced a visualization system \miscelaV{} for CAP mining and demonstrated the data analysis of smart city via \miscelaV.
We plan to continuously extend our system to improve usability and add additional data mining techniques, based on user feedback.
We hope that our system accelerates data analysis in many research fields.

\smallskip
\noindent
{\bf{Acknowledgements}} This work was supported by JSPS KAKENHI Grant Numbers JP20H00584.

\bibliographystyle{ACM-Reference-Format}
\bibliography{sample-base}

%
% ...

\end{document}